\documentclass[aps,prl,article,twocolumn,preprintnumbers,amsmath,amssymb,superscriptaddress,longbibliography]{revtex4-2}

\usepackage{amsmath}
\usepackage{amssymb}
\usepackage{bm}
\usepackage{graphicx}
\usepackage{xcolor}
\usepackage{booktabs}
\usepackage[colorlinks=true,linkcolor=blue,citecolor=blue,urlcolor=blue]{hyperref}

\newcommand{\pd}{P_d}

\newcommand{\figSmodel}{S1}

\newcommand{\figStcuthree}{S6}
\newcommand{\figSsigma}{S10}

\newcommand{\figStfour}{S5}
\newcommand{\figSkappa}{S17}
\newcommand{\figSfreq}{S7}

\newcommand{\figSdopten}{S11}

\newcommand{\figStstar}{S18}
\newcommand{\figSdmrg}{S19}

\begin{document}

\title{Phonon Fluctuations Enhanced Superconductivity in a Three-Band Cuprate Model}

\author{Jiarui Liu}
\email{jry\_liu@berkeley.edu}
\affiliation{Department of Physics, University of California, Berkeley, California 94720, USA}
\author{Yao Wang}
\email{yao.wang@emory.edu}
\affiliation{Department of Chemistry, Emory University, Atlanta, Georgia 30322, USA}
\begin{abstract}
The microscopic origin of high-$T_c$ superconductivity in cuprates remains an outstanding puzzle due to the interplay of intertwined orders. Experimental observations and theoretical studies have suggested crucial roles of phonons, with the $B_{1g}$ oxygen buckling mode emerging as a promising candidate. Here, we investigate the microscopic influence of this phonon in a three-band cuprate model. We find that the $B_{1g}$ phonon plays opposite roles in superconductivity depending on whether it lies in the adiabatic or antiadiabatic regime at a given temperature. Thermally populated adiabatic phonons suppress $d$-wave pairing through the loss of quasiparticle coherence, whereas antiadiabatic phonons simultaneously enhance coherence and the attractive $d$-wave pairing vertex, strengthening superconductivity across multiple diagnostics. This enhancement also exhibits pronounced electron-hole asymmetry arising from the different orbital characters of doped carriers. Our results therefore establish a microscopic mechanism through which $B_{1g}$ phonons can cooperate with electronic correlations to substantially enhance $d$-wave superconducting pairing in cuprates.
\end{abstract}

\maketitle

Understanding the origin of high-temperature $d$-wave superconductivity in doped cuprates remains a central challenge in condensed matter physics\,\cite{bednorz1986possible,shen1993anomalously,hardy1993precision,wollman1993experimental,tsuei1994pairing}, due to multiple intertwined orders at comparable length and energy scales\,\cite{fradkin2015colloquium}. Comparative studies of different cuprate families and other transition-metal oxides with Mott-like insulating parent compounds have highlighted the importance of charge-transfer physics and the role of oxygen ligands\,\cite{ruan2016relationship, weber2012scaling}. From a numerical perspective, the Hubbard and Emery models successfully capture the Mott insulating state, antiferromagnetism and several competing orders\,\cite{hubbard1963electron, emery1987theory, singh2002spin, huang2017numerical, zheng2017stripe, huang2023fluctuating}, but have not yet established a broadly robust description of hole-doped superconductivity\,\cite{jiang2019superconductivity, qin2020absence, chung2020plaquette, jiang2021ground, jiang2024ground, xu2024coexistence, yang2024recovery, zhao2025enhanced, jiang2026eightleg}. Recent comparisons between calculations and spectroscopy in quasi-one-dimensional cuprates point to an additional near-neighbor attraction beyond the minimal Hubbard description\,\cite{chen2021anomalously, shen2024signatures, li2025doping}, possibly mediated by phonons\,
\cite{wang2021phonon,tang2023traces}. These developments motivate the central question of whether realistic lattice vibrations can provide an additional pairing interaction that reinforces superconductivity in strongly correlated cuprates. 

Among the low-energy lattice modes observed in cuprates, the $B_{1g}$ oxygen buckling phonon is a particularly promising candidate. Its coupling to electronic states is strongly concentrated near the antinodal regions and carries an $x^2-y^2$ form factor, making it closely connected to the $d$-wave pairing symmetry\,\cite{devereaux2004anisotropic,cuk2004coupling,honerkamp2007phonons,macridin2012suppression}. Experimentally, angle-resolved photoemission measurements on Bi-2212 have revealed correlated enhancements of the superconducting gap, transition temperature, and electron--phonon coupling across the pseudogap boundary\,\cite{he2018rapid}, consistent with phonon anomalies observed in Raman and neutron scattering measurements\,\cite{pyka1993superconductivity, oh2021b}. Further evidence comes from the systematic increase of $T_c$ from single-layer to multilayer Ruddlesden--Popper compounds within the same material family. This layer dependence can be phenomenologically associated with the $B_{1g}$ electron--phonon coupling, which is forbidden by symmetry when the CuO$_2$ plane lies on a mirror plane but becomes allowed once this symmetry constraint is removed\,\cite{opel1999physical,johnston2010systematic}. Together, these observations suggest a close phenomenological connection between the $B_{1g}$ phonon and superconductivity, although the underlying microscopic mechanism remains unclear.

Here, we investigate the microscopic role of the $B_{1g}$ phonon in cuprate superconductivity using finite-temperature determinant quantum Monte Carlo (DQMC) simulations of a three-band Emery model, complemented by ground-state density-matrix renormalization group (DMRG) calculations. We reveal that its contribution to superconductivity changes qualitatively depending on the ratio between its frequency ($\Omega$) and temperature ($T$). At temperatures much higher than the phonon frequency, thermally populated phonons suppress quasiparticle coherence, leading to weakened superconducting correlations. By contrast, in the antiadiabatic regime, whose numerical criterion is consistent with cuprate experiments, the $B_{1g}$ phonon cooperatively enhances coherence and the $d$-wave pairing vertex, resulting in a substantial enhancement of superconducting pairing. This enhancement is particularly pronounced on the underdoped side of the superconducting dome and exhibits a significant electron-hole asymmetry. Complementary ground-state calculations further demonstrate that antiadiabatic phonons enhance $d$-wave pairing correlations while suppressing the competing stripe order. These results reveal the microscopic mechanism through which $B_{1g}$ phonons enhance $d$-wave superconductivity in cuprates.

\begin{figure*}[!t]
\centering
\includegraphics[width=\textwidth]{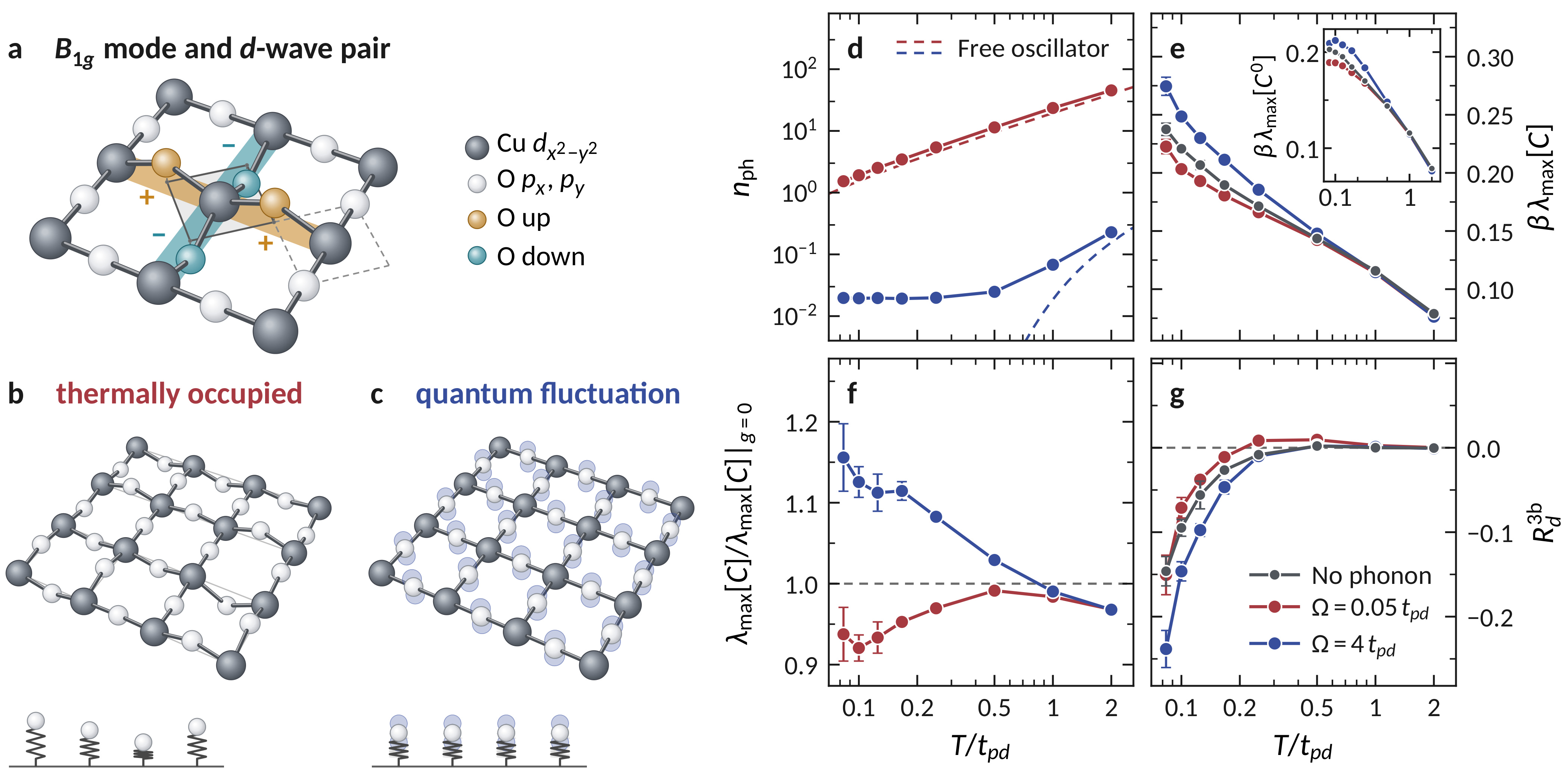}\vspace{-3mm}
\caption{\label{fig:temperature}\textbf{Temperature-dependent effects of $B_{1g}$ phonons on superconducting pairing.} \textbf{a}, Schematic illustration of the $B_{1g}$ oxygen buckling mode, where oxygen atoms along the $x$ and $y$ directions of a plaquette move in opposite out-of-plane directions. The dashed box indicates a $d$-wave pair in the CuO$_2$ plane, whose order parameter changes sign under a 90$^\circ$ rotation. \textbf{b},\textbf{c}, Schematic illustrations of an adiabatic phonon mode in \textbf{b}, which is thermally populated and exhibits large quasi-static distortions (inset), and an antiadiabatic phonon mode in \textbf{c}, which is dominated by quantum fluctuations around the equilibrium positions. \textbf{d}, DQMC results for the average phonon occupation versus temperature in the $8\times8$ Emery model at $p=1/8$ doping, coupled to the adiabatic mode (red, $\Omega=0.05t_{\rm pd}$) and antiadiabatic mode (blue, $\Omega=4t_{\rm pd}$). The dashed lines indicate the Bose occupation $1/(e^{\Omega/T}-1)$ of a free phonon of the same frequency. \textbf{e}, Three-orbital midpoint pairing correlation $\beta\lambda_{\max}[C]$ for the same phonon-coupled models as in \textbf{d}, compared with the purely electronic model (gray). The inset shows the temperature dependence of the dressed bubble $\beta\lambda_{\max}[C^0]$. \textbf{f}, Relative changes in the pairing correlations shown in \textbf{e}, normalized to the purely electronic model. \textbf{g}, Pairing vertex ratio $R_d^{\mathrm{3b}}$ for the purely electronic Emery model (gray) and the models coupled to adiabatic (red) and antiadiabatic (blue) phonons. Error bars represent one standard error in all panels.}
\end{figure*}

\section{The Three-Band Model Coupled to a $B_{1g}$ Buckling Phonon}

We consider the three-band Emery Hamiltonian containing one Cu $d_{x^2-y^2}$ orbital and two O $p_{x,y}$ orbitals per unit cell, with a local phonon assigned to each CuO$_4$ plaquette. The full Hamiltonian takes the form
\begin{equation}\label{eq:model}
    \mathcal{H}=\mathcal{H}_{e}+\Omega\sum_R \hat b_R^\dagger\hat b_R+g\sum_R(\hat b_R+\hat b_R^\dagger)\hat B_R\,,
\end{equation}
where the purely electronic part is described by
\begin{align}
    \mathcal{H}_{e}={}&-\sum_{\langle ij\rangle\sigma}t^{ij}_{pd}\big(\hat d^\dagger_{i\sigma}\hat p_{j\sigma}+\mathrm{h.c.}\big)-\sum_{\langle jj'\rangle\sigma}t^{jj'}_{pp}\big(\hat p^\dagger_{j\sigma}\hat p_{j'\sigma}+\mathrm{h.c.}\big)\nonumber\\
    &+(\epsilon_p-\epsilon_d)\sum_{j\sigma}\hat n^{p}_{j\sigma}+U_d\sum_i\hat n^d_{i\uparrow}\hat n^d_{i\downarrow}\,,\label{eq:emery}
\end{align}
with hopping amplitudes $t^{ij}_{pd}=\pm t_{\rm pd}$ and $t^{jj'}_{pp}=\pm t_{pp}$ whose signs follow the corresponding orbital phases (SM, Sec.~S1). The electron--phonon coupling term in Eq.~\eqref{eq:model} describes the interaction between the $B_{1g}$ oxygen buckling mode and the local charge distribution, characterized by the operator
\begin{equation}\label{eq:br}
    \hat B_R=\hat n^x_R+\hat n^x_{R-\hat x}-\hat n^y_R-\hat n^y_{R-\hat y}\,,
\end{equation}
where the opposite signs of the $p_x$ and $p_y$ oxygen densities reflect the $B_{1g}$ symmetry around the Cu site $R$ (Fig.~\ref{fig:temperature}a and Fig.~\figSmodel a). Following previous studies, we adopt $t_{\rm pd}=1.13$\,eV, $t_{pp}=0.49$\,eV, $\epsilon_p-\epsilon_d=3$\,eV, and $U_d=6$\,eV\,\cite{huang2017numerical}. Unless otherwise specified, we fix the electron--phonon coupling strength at $\kappa=g^2/\Omega=0.25t_{\rm pd}$ to maintain numerical control and consistency across the investigated parameter range.

The $B_{1g}$ oxygen buckling mode exhibits a characteristic $d$-wave-like displacement pattern\,\cite{devereaux2004anisotropic,cuk2004coupling,honerkamp2007phonons,macridin2012suppression}, in which the two $p_x$ oxygens move out of the CuO$_2$ plane in the opposite direction to the two $p_y$ oxygens within each plaquette (Fig.~\ref{fig:temperature}a). In the adiabatic limit (Fig.~\ref{fig:temperature}b), where $T\gg \Omega$, thermally populated phonons primarily influence the electrons through the average lattice displacement described in Eq.~\eqref{eq:model}. By contrast, in the antiadiabatic limit (Fig.~\ref{fig:temperature}c), where $T\ll \Omega$, quasi-static lattice distortions are strongly suppressed. The phonon then influence the electronic system through quantum fluctuations.

To quantify superconducting pairing, we examine both the Cu-projected pairing correlation and the correlations among four orthonormal $B_{1g}$ pair operators constructed from the three orbitals (SM, Sec.~S4). We first define the dynamical Cu--Cu $d$-wave pairing correlator $\pd(\tau)$ as
\begin{equation}\label{eq:pd}
\begin{split}
    \pd(\tau)={}&\frac1{4N}\sum_{R,R'}\sum_{\delta,\delta'}f_\delta f_{\delta'}\\
    &\times\big\langle \hat d_{R\downarrow}(\tau)\hat d_{R+\delta\uparrow}(\tau)\hat d_{R'+\delta'\uparrow}^\dagger(0)\hat d_{R'\downarrow}^\dagger(0)\big\rangle\,,
\end{split}
\end{equation}
where $f_{\pm\hat x}=1$ and $f_{\pm\hat y}=-1$ define the $d$-wave form factor, and $N$ denotes the number of Cu sites. Evaluating the correlation at the imaginary-time midpoint, $P_d(\beta/2)$ with $\beta=1/T$, suppresses spectral contributions at $|\omega|\gg T$ and isolates the low-energy electronic scale relevant to the emerging pairing instability\,\cite{wang2025finite}. We further generalize the pairing correlation in Eq.~\eqref{eq:pd} to a $4\times4$ matrix $C_{mn}(\tau)$ constructed from these four pair operators. The dominant pairing response is then characterized by its largest eigenvalue $\lambda_{\max}[C]$\,\cite{zhao2025enhanced}. 

In addition to the pairing correlations, we examine the pairing vertex through the ratio
\begin{equation}\label{eq:rd3b}
    R_d^{\mathrm{3b}}=\frac{\lambda_{\max}[C^0]}{\lambda_{\max}[C]}-1\,,
\end{equation}
where $C^0$ denotes the bubble contribution to $C$, constructed from the interacting single-particle Green functions. The vertex ratio $R_d^{\mathrm{3b}}$ becomes negative when the full pairing response exceeds the bubble contribution\,\cite{white1989attractive, wang2025finite}, with increasingly negative values indicating a stronger attractive $d$-wave pairing vertex. This decomposition therefore allows us to distinguish changes in low-energy electronic propagation from genuine vertex contributions to superconducting pairing.

\section{Opposite Roles of Phonons at the Two Limits}

At fixed $1/8$ doping on an $8\times 8$ cluster, we examine the effects of $B_{1g}$ phonons over a temperature range from $T=t_{\rm pd}/12$ to $2t_{\rm pd}$.  We first consider a $B_{1g}$ phonon with $\Omega=0.05t_{\rm pd}$, representing the adiabatic regime in which thermal phonon occupation is substantial and increases with temperature (Fig.~\ref{fig:temperature}d). By tracking the pair-field susceptibility $\lambda_{\max}[C]$, we find that the $B_{1g}$ phonon suppresses $d$-wave pairing compared with the bare three-band Emery model. As quantified by the relative suppression in Fig.~\ref{fig:temperature}f, this effect is most pronounced at $T<0.5t_{\rm pd}$.  As the temperature increases further, the intrinsic electronic pairing correlations of the Emery model become progressively weaker, making the additional suppression by the phonon less distinguishable. 

By contrast, when considering a $B_{1g}$ phonon with $\Omega=4t_{\rm pd}$ as a representative high-frequency mode in the antiadiabatic regime, we observe the opposite trend. The phonon occupation remains small in this regime (Fig.~\ref{fig:temperature}d). As shown in Figs.~\ref{fig:temperature}e and f, the high-frequency phonon enhances $d$-wave pairing at all temperatures below $t_{\rm pd}$, with the enhancement increasing continuously toward the lowest temperature accessible in our simulations. As the temperature increases above $t_{\rm pd}$, the low-temperature enhancement gradually gives way to a suppression of pairing (Fig.~\ref{fig:temperature}f), indicating a crossover toward thermally dominated phonon effects. 

We emphasize that the adiabatic or antiadiabatic behavior of a phonon depends not only on its absolute frequency but also on the relative temperature scale. At sufficiently high temperatures, thermal fluctuations become important even for the $\Omega=4t_{\rm pd}$ phonon, causing the pair-field susceptibilities of the two phonon-frequency regimes to converge (Figs.~\ref{fig:temperature}e and f). Conversely, at the lowest temperature $T=t_{\rm pd}/12$, the pair-field susceptibility associated with the $\Omega=0.05t_{\rm pd}$ phonon already shows a tendency toward enhancement, although the accessible temperature range in our simualtions does not allow us to establish this behavior conclusively

The phonon-induced enhancement also increases monotonically with the electron--phonon coupling strength, as demonstrated at $T=0.125t_{\rm pd}$ (Fig.~\figSkappa). Importantly, in both the adiabatic and antiadiabatic regimes, the Cu--Cu pairing correlation $\pd$ follows the same trend as the multiorbital susceptibility $\lambda_{\max}[C]$ (Fig.~\figStcuthree), indicating that the Zhang--Rice-type dopant wave function remains preserved at this doping\,\cite{zhang1988effective}.

\section{Contributions to Coherence and Vertex}

To identify the microscopic mechanism underlying the phonons' impacts on pairing, we decompose the pair-field susceptibility into the dressed bubble and pairing vertex, which separately characterize the contributions from quasiparticle coherence and effective pairing interactions\,\cite{macridin2012suppression}. At the lowest temperature ($T=t_{\rm pd}/12$), the adiabatic phonon substantially reduces the dressed bubble $\lambda_{\max}[C^0]$, while leaving the vertex ratio $R_d^{\mathrm{3b}}$ nearly unchanged relative to the purely electronic model (Fig.~\ref{fig:temperature}e,g). The resulting suppression of pairing therefore originates primarily from the loss of low-energy coherence caused by incoherent scattering from thermally populated phonons. Consistent with this interpretation, the adiabatic phonon lowers the low-energy Cu spectral weight $G(\bm k,\beta/2)$ at both the antinode and the node, increasingly on cooling (Fig.~\figSsigma). In sharp contrast, the antiadiabatic phonon elevates both the dressed bubble and the attractive pairing vertex, allowing these two contributions to cooperatively enhance superconducting correlations. Remarkably, at the lowest temperature, it makes the vertex ratio $R_d^{\mathrm{3b}}$ about 60\% more negative than in the bare electronic model (Fig.~\ref{fig:temperature}e,g and Fig.~\figStfour).

\begin{figure}[tb]
\centering
\includegraphics[width=\columnwidth]{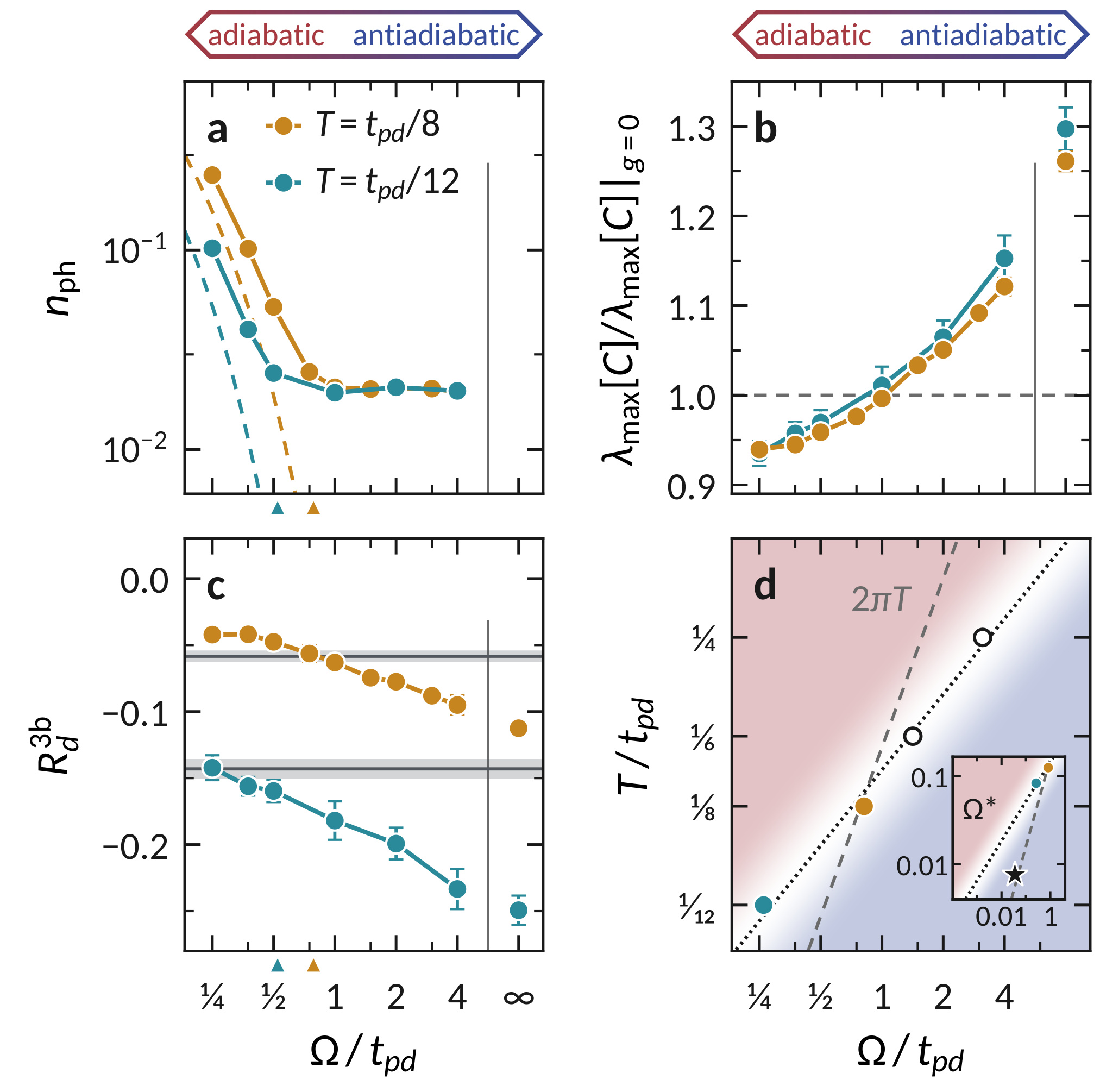}\vspace{-3mm}
\caption{\label{fig:frequency}\textbf{Frequency dependence of the $B_{1g}$ phonon effect on $d$-wave pairing.} \textbf{a}, Average phonon occupation $n_{\mathrm{ph}}$ as a function of phonon frequency. Results are calculated at $1/8$ doping for $T=t_{\rm pd}/8$ and $t_{\rm pd}/12$. The dashed curves indicate the Bose occupations of independent phonons, while the bottom triangles mark $\Omega=2\pi T$ for each temperature. \textbf{b}, Relative enhancement or suppression of the pairing susceptibility $\lambda_{\max}[C]$ induced by phonons, compared with the purely electronic model, as a function of phonon frequency. The dashed line indicates unity. \textbf{c}, Pairing vertex ratio $R_d^{\mathrm{3b}}$ as a function of phonon frequency at the same two temperatures. The gray bars indicate the corresponding vertex ratios of the purely electronic model, with their half-widths representing one standard error. \textbf{d}, Critical frequency $\Omega^*(T)$ above which the pairing vertex ratio $R_d^{\mathrm{3b}}$ falls below that of the purely electronic model. Open symbols indicate $T=t_{\rm pd}/4$ and $t_{\rm pd}/6$, not shown in \textbf{a}--\textbf{c}. The dotted line is a power-law fit to the four symbols and the dashed line indicates $\Omega=2\pi T$. The red and blue shading marks the phonon-induced vertex suppression and enhancement. The inset indicates the position of the cuprate's $B_{1g}$ phonon ($40$\,meV, star) at $T=100$\,K in the extended diagram.}

\end{figure}

The rapid low-temperature growth of the pairing vertex allows us to estimate the characteristic temperature through extrapolation. We fit the vertex with $R_d^{\mathrm{3b}}=a+b/T$ and define $T^*$ by the extrapolated condition $R_d^{\mathrm{3b}}=-1$ (Fig.~\figStstar), indicating the divergence of pair-field susceptibility. By applying the same estimator and fitting procedure to all simulated systems, we use $T^*$ as a consistent proxy for the superconducting transition temperature. The inclusion of the antiadiabatic $B_{1g}$ phonon increases $T^*$ by about 34\%, with the bootstrap interval for the enhancement ratio lying entirely above unity.

To further confirm the mechanism underlying the suppression and enhancement of superconductivity, we systematically vary the phonon frequency while keeping the temperature and coupling strength $\kappa$ fixed. As shown in Fig.~\ref{fig:frequency}a, the phonon occupation generally decreases with increasing phonon frequency. In particular, it drops sharply between $\Omega=0.25t_{\rm pd}$ and $t_{\rm pd}$, before approaching saturation when $\Omega\gtrsim 5 T$, marking the crossover from the thermally occupied to the quantum-fluctuation-dominated regime. Remarkably, this crossover approximately coincides with the transition of the pair-field susceptibility from suppression to enhancement relative to the purely electronic model (Fig.~\ref{fig:frequency}b).

Similarly, we examine the phonon-frequency dependence of quasiparticle coherence and the pairing vertex. Of these two contributions, the pairing vertex plays a crucial role in determining the overall effect of phonons on superconducting pairing. For example, at $T=t_{\rm pd}/12$, the dressed bubble remains below its phonon-free value up to $\Omega\approx t_{\rm pd}$ and returns to it only near $2t_{\rm pd}$ (Fig.~\figSfreq). Nevertheless, this suppression is compensated by the enhancement of the attractive pairing vertex (Fig.~\ref{fig:frequency}c and SM, Sec.~S9), so that near $2t_{\rm pd}$ the pair-field susceptibility is already enhanced while the bubble is not (Fig.~\ref{fig:frequency}b).

The vertex gain requires the phonon to remain dynamical on the thermal energy scale. Analytically, the phonon propagator takes the form $2\Omega/(\nu_n^2+\Omega^2)$ at bosonic Matsubara frequencies $\nu_n=2\pi nT$. When $\Omega\ll2\pi T$ (adiabatic), its contributions at all nonzero $\nu_n$s become negligible, and the thermally occupied phonon acts primarily as a quasi-static distortion. As $\Omega$ increases beyond the thermal energy scale, however, the phonon can mediate an effective attractive interaction. Correspondingly, the phonon-induced pairing vertex becomes appreciably more attractive when $\Omega$ approaches or exceeds $2\pi T$.  This mechanism becomes particularly transparent in the extreme antiadiabatic limit ($\Omega\to\infty$), where the phonon can be integrated out and Eq.~\eqref{eq:model} reduces to the instantaneous interaction $-\kappa\sum_R\hat B_R^2$ (SM, Sec.~S1). Projecting this interaction onto the Zhang--Rice-like Cu--O hybrid states\,\cite{zhang1988effective}, yields an explicit attractive interaction in the $d$-wave pairing channel (SM, Sec.~S2),
\begin{equation}\label{eq:hinf}
    -\frac{4\lambda_{1\mathrm b}t_{\mathrm{eff}}}{N}\sum_{kk'}(\cos k_x\!-\cos k_y)(\cos k_x'\!-\cos k_y')\hat\Delta^\dagger_{k'}\hat\Delta_k\,,
\end{equation}
where $\hat\Delta_k$ annihilates a Cooper pair in the hybrid state, and the dimensionless coupling $\lambda_{1\mathrm b}$ is proportional to $\kappa$ (SM, Sec.~S2).

Interestingly, in our simulations with strong electronic interactions, the numerical crossover extends to higher temperatures than suggested by this simple $\Omega\sim 2\pi T$ criterion (Fig.~\ref{fig:frequency}d). This suggests that the conditions for phonon-enhanced pairing may be more favorable in strongly correlated cuprates, particularly in the experimentally relevant 100\,K regime.

\begin{figure}[tb]
\centering
\includegraphics[width=\columnwidth]{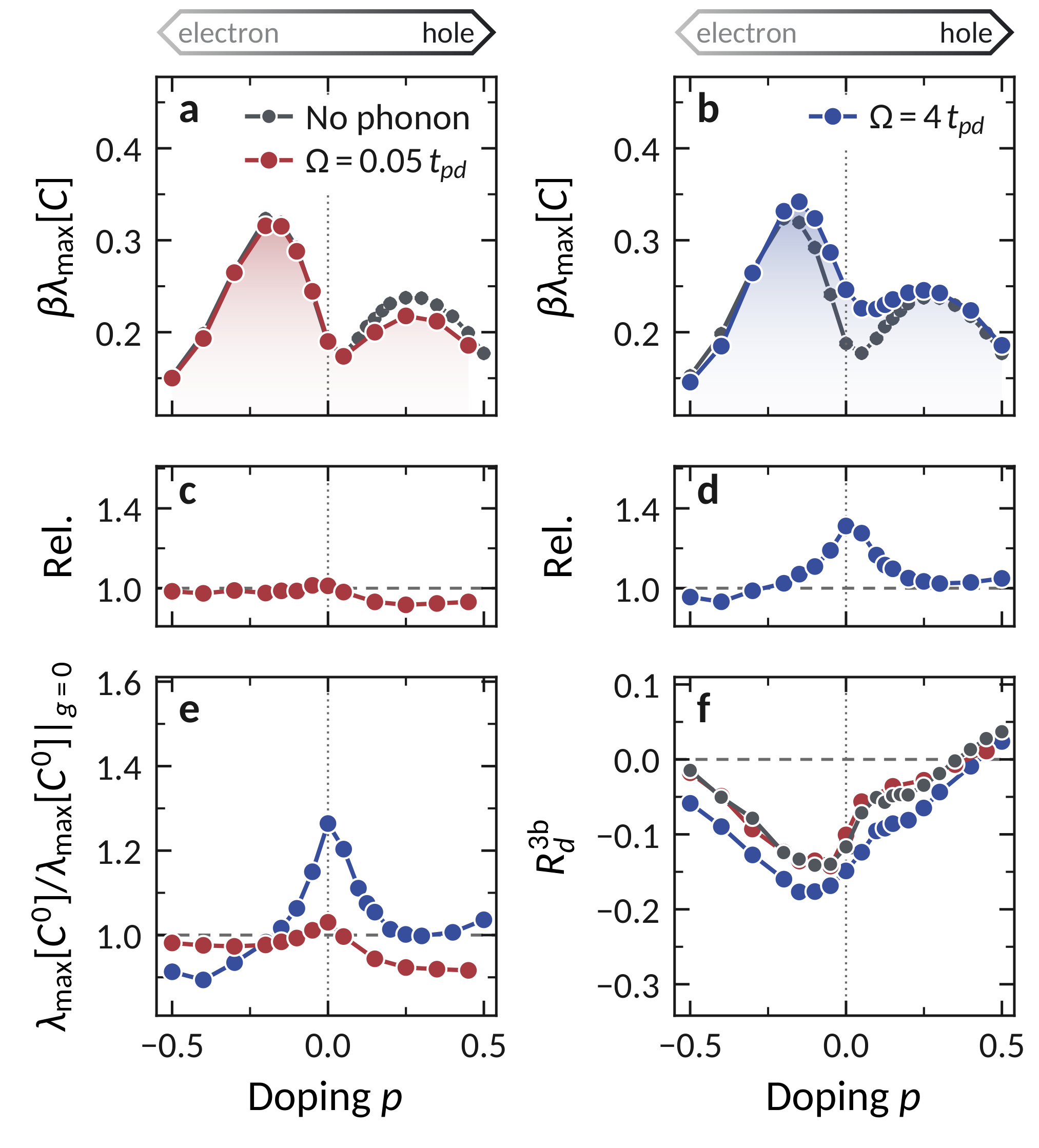}
\vspace{-5mm}
\caption{\label{fig:doping}\textbf{Doping dependence of the $B_{1g}$ phonon effects.} \textbf{a},\textbf{b}, Doping dependence of the midpoint pairing correlation $\beta\lambda_{\max}[C]$ for the purely electronic model (gray) and the models coupled to adiabatic phonons (red, $\Omega=0.05t_{\rm pd}$) in \textbf{a} and antiadiabatic phonons (blue, $\Omega=4t_{\rm pd}$) in \textbf{b}. \textbf{c},\textbf{d}, Relative enhancement or suppression of the pairing susceptibility $\lambda_{\max}[C]$ induced by the adiabatic phonon in \textbf{c} and the antiadiabatic phonon in \textbf{d}, normalized to the purely electronic model. \textbf{e}, The same ratio for the dressed bubble $\lambda_{\max}[C^0]$ for both phonons. The dashed lines indicate unity, corresponding to the purely electronic model. \textbf{f}, Doping dependence of the pairing vertex ratio $R_d^{\mathrm{3b}}$ for the same models, using the color notation in \textbf{a} and \textbf{b}. All results are obtained at temperature $T=t_{\rm pd}/8$. Error bars represent one standard error. The same quantities at $T=t_{\rm pd}/10$ are shown in Fig.~\figSdopten.}
\end{figure}

\section{Doping Dependence}

The contrasting effects of adiabatic and antiadiabatic phonons on superconducting pairing persist throughout the hole-doped regime, although their magnitudes vary with doping. As shown in Fig.~\ref{fig:doping}a,b, the pairing susceptibility of the purely electronic model exhibits a superconducting dome centered near 25\% hole doping (gray curve). The inclusion of the adiabatic $B_{1g}$ phonon substantially suppresses this dome (Fig.~\ref{fig:doping}a,c). This suppression primarily originates from the reduction of quasiparticle coherence, which becomes substantial away from half-filling (Fig.~\ref{fig:doping}e), while the pairing vertex remains nearly unchanged (Fig.~\ref{fig:doping}f). By contrast, the antiadiabatic phonon makes the pairing vertex more attractive at every hole doping, most strongly at low doping (Fig.~\ref{fig:doping}f). Its enhancement of quasiparticle coherence is largest near half-filling and nearly vanishes for $p\simeq0.2$--0.3 (Fig.~\ref{fig:doping}e), while the overall pairing susceptibility exhibits a maximum relative enhancement of approximately 30\% toward the low-hole-doping side (Fig.~\ref{fig:doping}d). This enhancement becomes weaker as doping approaches the Lifshitz transition of the Fermi surface near 30\% hole doping. The resulting doping dependence produces an asymmetric enhancement of the superconducting dome in Fig.~\ref{fig:doping}b, shifting its maximum toward lower hole doping.

The role of phonons changes substantially upon moving to the electron-doped regime. In the three-band description of cuprates, approximately two-thirds of the doped holes occupy O orbitals, whereas two-thirds of the doped electrons occupy Cu orbitals (SM, Sec.~S11)\,\cite{white2015doping, kung2016characterizing}. Because the $B_{1g}$ mode couples only to the oxygen charge density, the suppressive effects of adiabatic phonons become nearly negligible throughout the electron-doped regime, as reflected in the total pairing susceptibility, dressed bubble, and pairing vertex (Fig.~\ref{fig:doping}c,e,f). Meanwhile, although the antiadiabatic phonon continues to enhance the attractive pairing vertex nearly uniformly (Fig.~\ref{fig:doping}f), its contribution to the dressed bubble becomes more strongly doping dependent on the electron-doped side (Fig.~\ref{fig:doping}e). In particular, the enhancement of quasiparticle coherence is confined to a narrower doping window near half-filling, while the phonon instead suppresses the dressed bubble beyond 15\% electron doping. Consequently, the electron-hole asymmetry leads to a much weaker enhancement of $d$-wave pairing by the $B_{1g}$ phonon on the electron-doped side compared with the hole-doped side. This distinction may provide a microscopic contribution to understanding why superconductivity is generally more robust in hole-doped cuprates than in their electron-doped counterparts\,\cite{armitage2010progress}.

\section{Antiadiabatic Phonons at the Ground State}

\begin{figure}[!t]
\raggedright
\includegraphics[width=\columnwidth]{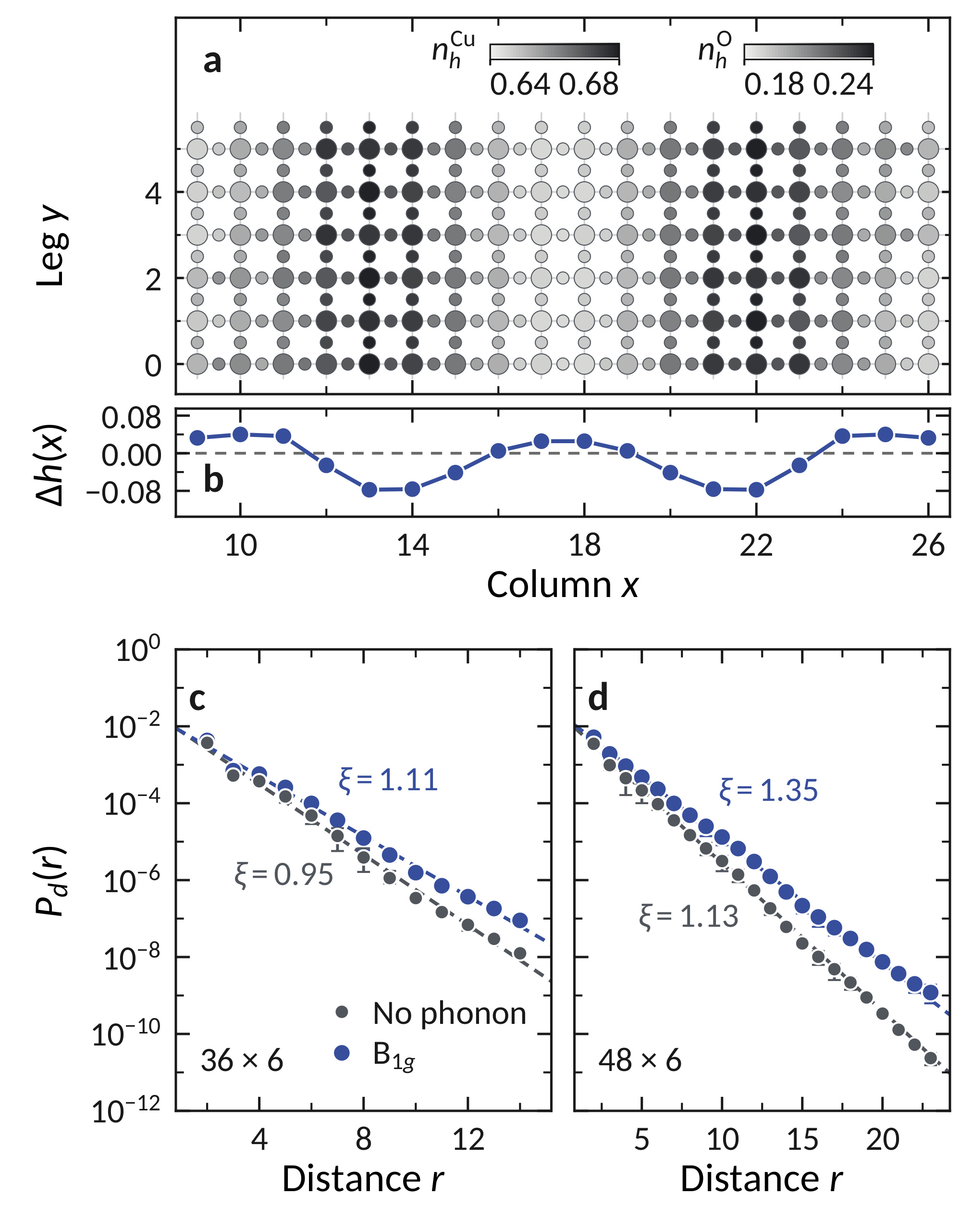}
\vspace{-5mm}
\caption{\label{fig:dmrg}\textbf{Charge stripes and pairing correlations in the ground state.} \textbf{a}, Hole-density distribution on Cu and O sites within the middle 18 columns of a $36\times6$ cylinder at hole doping $p=1/12$, coupled to an extremely antiadiabatic $B_{1g}$ phonon with $\Omega\to\infty$ and $\kappa=0.25t_{\rm pd}$. The results are obtained at the largest bond dimension reached, retaining $D=36\,000$ SU(2) multiplets. \textbf{b}, Difference in the column-resolved hole-density distribution, $\Delta h(x)$, relative to the purely electronic model, exhibiting a modulation opposite to the static hole-density distribution. \textbf{c},\textbf{d}, Static $d$-wave pairing correlation $P_d(r)$ on Cu sites as a function of distance $r$ for the $36\times6$ cylinder in \textbf{c} and the $48\times6$ cylinder in \textbf{d}, obtained by averaging over all legs and four reference rungs. The dashed lines indicate exponential fits to the pairing correlations. The truncation errors are $3.4\times 10^{-5}$ and $3.7\times 10^{-5}$, respectively.}
\end{figure}

To establish whether the pairing enhancement mechanism persists in the ground state, we perform complementary DMRG simulations in the $\Omega=\infty$ limit\,\cite{white1992density}, using the same coupling strength $\kappa=0.25t_{\rm pd}$. We compare the phonon-induced effective Hamiltonian with the bare three-band Emery model on $36\times6$ and $48\times6$ clusters at $p=1/12$. To ensure numerical accuracy, we retain up to $D=36\,000$ and $30\,000$ SU(2) multiplets in the sector symmetric under inversion of the cylinder (SM, Sec.~S15, and Fig.~\figSdmrg).

As shown in Fig.~\ref{fig:dmrg}a, the ground state of the bare Emery model develops pronounced stripe order, consistent with previous single-band calculations and three-band simulations\,\cite{white2015doping,zheng2017stripe,jiang2024ground,polat2025influence,zhang2026robust,jiang2026eightleg}. The inclusion of the antiadiabatic $B_{1g}$ phonon weakens this competing stripe order, although it remains present at the coupling strength considered here (Fig.~\ref{fig:dmrg}b). Meanwhile, the static $d$-wave pairing correlation $P_d(r)$ exhibits exponential spatial decay, indicating that the pairing fluctuations remain short-ranged on these cylinders at our best truncation. Nevertheless, the phonon substantially extends the spatial range of pairing correlations, increasing the extracted correlation length $\xi$ by 16--19\% at equal truncation error on the two cylinders. These results demonstrate that antiadiabatic $B_{1g}$ phonons strengthen static $d$-wave pairing correlations in the ground state, consistent with the enhanced dynamical pair-field susceptibility at finite temperature and the increased pairing scale $T^*$. This effect is also consistent with recent findings that near-neighbor attraction modifies the pairing-correlation power-law exponent of a four-leg single-band model or two-leg three-band model\,\cite{peng2023enhanced, jiang2023pair}, where the attractive interaction may also originate from phonons\,\cite{wang2021phonon, tang2023traces}.

\section{Discussion and Outlook}

Our results reveal that electronic and lattice mechanisms can cooperate to promote high-$T_c$ superconductivity. When an antiadiabatic phonon has negligible thermal occupation and contributes primarily through virtual processes, the resulting effective interaction makes the $d$-wave pairing vertex more attractive and, at sufficiently high frequencies, also improves low-energy electronic coherence. These cooperative effects enhance superconducting correlations across multiple metrics, including the pairing correlation length, effective pairing scale $T^*$, and pairing vertex. While these quantities are enhanced by approximately 20--60\% at the coupling strengths examined in our simulations, the enhancement generally increases with stronger electron--phonon coupling. Notably, the coupling $\kappa=0.25t_{\rm pd}$ adopted in our simulations is chosen to avoid orbital-polarization instabilities under all simulated conditions, but corresponds to a very small projected one-band coupling $\lambda_{1\mathrm b}\approx0.02$ in the form of Eq.~\eqref{eq:hinf}, where the orbital degrees of freedom are frozen (SM, Sec.~S2).

The contrasting effects of the $B_{1g}$ phonon imply that its frequency and thermal occupation, rather than its symmetry alone, determine whether phonons enhance or suppress pairing, making the phonon energy an important parameter for controlling superconductivity. Although our DQMC simulations require artificially high phonon frequencies to reach the antiadiabatic regime at accessible temperatures, the typical experimental temperature is $T_c\simeq100$\,K and the cuprate $B_{1g}$ phonon energy is approximately 36--42\,meV\,\cite{pyka1993superconductivity,oh2021b,he2018rapid,cuk2004coupling}. This phonon energy is comparable to the analytical criterion $\Omega\sim2\pi T$. Importantly, as illustrated in Fig.~\ref{fig:frequency}d, strong electronic correlations reduce the crossover energy scale relative to this simple criterion, placing the physical $B_{1g}$ phonon deeply into the antiadiabatic regime (Fig.~\ref{fig:frequency}), cooperatively supporting superconductivity. 

\bibliography{main_arxiv}

\section*{Acknowledgments}
We thank Joseph Checkelsky, Thomas Devereaux, Ruihua Fan, Yu He, Manki Kim, Steven Kivelson, Grace Pan, and Zhi-Xun Shen for insightful discussions. This work is primarily supported by the Periodic Labs research grant.

\section*{Author contributions:} 
Y.W. conceived the idea of the project. J.L. developed code and performed numerical simulations. J.L. and Y.W. wrote the manuscript together. 

\section{Competing interests}
The authors declare no competing interests.

\end{document}